\documentstyle[eqsecnum,aps,preprint,tighten]{revtex}
\begin{document}
\draft
\def\beq{\begin{eqnarray}}
\def\eeq{\end{eqnarray}}
\def\nn{\nonumber\\}

\preprint{\vbox{\baselineskip=12pt
\rightline{CERN-TH/98-199}
\rightline{}
\rightline{hep-th/9806233}}}
\title{Quantum Fluctuations in Open Pre-Big Bang  Cosmology}
\author{A. Ghosh, G. Pollifrone and  G. Veneziano}
\address{Theory Division, CERN, 1211 Geneva 23, Switzerland}
\date{\today}
\maketitle
\begin{abstract}
We solve exactly the (linear order) equations
for tensor and scalar perturbations over the homogeneous, isotropic,
 open  pre-big bang model
 recently discussed by several authors. We find that the
parametric amplification of vacuum fluctuations (i.e. particle
production) remains negligible throughout
the perturbative pre-big bang phase.
\end{abstract}
\pacs{\tt PACS number(s):~98.80.Cq,~04.30.Db}

\section{Introduction}
The question
of whether, in the presence of spatial curvature,
the pre-big bang (PBB) scenario
\cite{PBB,veneziaerice,webpage} needs a very large amount of
 fine-tuning is still a subject
of debate \cite{inhomog,TW,maggiore,BMUV,KLB,MOV,BDV}.
Furthermore, Kaloper et al. \cite{KLB} have argued that, 
even  assuming that
 the two classical moduli of the open (${\cal K}=-1$), homogeneous, isotropic
 cosmological solution \cite{CLW,TW} lie
deeply inside the perturbative region, the unavoidable existence of vacuum
quantum fluctuations
modifies so drastically  the classical behaviour as to prevent the 
occurrence of an appreciable amount of inflation.

In this paper, as a first step towards addressing this second
objection, we  carry out a detailed study
 of  quantum fluctuations around the ${\cal K}=-1$ solution
 of \cite{CLW,TW}. It is well known \cite{MFB,GMST} that
 quantum fluctuations
in a non-spatially flat background are considerably harder to study  
than the corresponding ones in a flat Universe.
Nevertheless, somewhat to our surprise,
the corresponding equations can still be integrated {\it exactly}
in terms of standard hypergeometric functions.
The conclusion is that particle production (i.e. the amplification
of vacuum fluctuations) is strongly suppressed at very early times
because of a 
cancellation between the effect of a non-vanishing Hubble parameter
and the one of spatial curvature. In other words, particle production
is proportional to the deviation of the background from its asymptotic
Milne form and thus  to the time variation of the background dilaton.
As a result, particle production 
remains small through the whole perturbative
PBB phase and does not impede the occurrence of PBB inflation.

We will first recall the explicit form of the homogeneous, 
isotropic, ${\cal K}=-1$
PBB background we shall be dealing with
and derive the general, covariant form of the action to second order  
in the perturbations.  We then solve, successively,
 the equations for tensor and scalar perturbations. Finally,
we discuss the physical implications of our results, and comment
on their possible relevance to the issue raised in  \cite{KLB}.

\section{The background and the second-order action}

Our conventions are such that (after reduction to $D=4$)
the (normalized) string-frame action   takes
the form
\beq
\hbar^{-1} S^{(s)} ={1\over 2\ell_s^2}\int d^4x\sqrt{-G}e^{-\phi}
(R(G)+G^{\mu\nu}\partial_\mu\phi
\partial_\nu\phi +\dots)\; ,
\label{Saction}
\eeq
where $G_{\mu\nu}$ is the string-frame metric, $\phi$ is the ($D=4$)  
dilaton, $\ell_s$ is the fundamental
length scale of string theory and the dots indicate other fields  
(e.g. a Kalb-Ramond axion field) that will be set to zero hereafter.
 The above action allows for classical  homogeneous, isotropic
 solutions of the standard Friedmann--Robertson--Walker
(FRW) type
\beq
ds^2=a^2_s(\eta)\left(-d\eta^2+{dr^2\over 1-{\cal K}r^2}+r^2d\Omega^2\right).
\eeq
As usual there are both post- and pre-big bang solutions coming from  
a singularity,
or going towards it, respectively.
For ${\cal K}=-1$, the PBB-type solution was first given in \cite{CLW} and
then rederived and
discussed in \cite{TW}. It reads:
\beq
a_s(\eta)&=& L (\cosh\eta)^{1+\sqrt 3\over 2}
(-\sinh\eta)^{1-\sqrt 3\over 2}\nn
\phi(\eta)&=&-\sqrt 3\ln(-\tanh\eta)+\phi_{\text{in}} \; , \;\: \eta  
< 0 \; ,
\label{Ssol}
\eeq
where $L$ and $\phi_{\text{in}}$ are a dimensional and a  
dimensionless integration constant, respectively.

The arbitrariness of $L$ and $\phi_{\text{in}}$ reflects the
 symmetries of the classical problem under a constant shift of the
dilaton $\phi$ and a constant rescaling of the metric $G_{\mu\nu}$.
These are precisely the two parameters
to be chosen in an appropriate (fine-tuned 
\cite{inhomog,TW,maggiore,BMUV,KLB,MOV,BDV}?) range in order to ensure
a sufficient amount of PBB inflation. Indeed, Eq. (\ref{Ssol}) describes
a universe that is almost trivial (Milne-like) from $-\infty < \eta <
{\cal O} (-1)$,
and then inflates with an initial curvature ${\cal O}(L^{-2})$ and initial
coupling ${\cal O}(\exp(\phi_{\text{in}}/2))$ till it meets, eventually,
 the strong curvature and/or strong coupling
regimes at $\eta \sim \eta_1$. The critical value $\eta_1$ is easily
 determined in terms of
 the integration constants $L$ and $\phi_{\text{in}}$:
\beq
(- \eta_1) = 
\max\,( e^{\phi_{\text{in}}/\sqrt{3}} , (\ell_s/L)^{1 + 1/\sqrt{3}} )\; .
\eeq

It is well known \cite{PBB} that the study of
perturbations is technically  simpler in the so-called Einstein  
frame, defined by $g_{\mu\nu}=\exp(\phi_{\text{today}} -\phi)  
G_{\mu\nu}$, and,
correspondingly, by the action:
\beq
\hbar^{-1}S^{(E)}={1\over 2\ell_P^2}\int d^4x\sqrt{-g}\left(R(g) -{1\over 2}
g^{\mu\nu}\partial_\mu\phi
\partial_\nu\phi\right)\; ,
\label{Eaction}
\eeq
where $\phi_{\text{today}}$ is the present value of the dilaton and
$\ell_P \equiv \sqrt{8 \pi G \hbar} = \exp(\phi_{\text{today}}/2)  
\ell_s \sim 0.1 \ell_s$
is the present value of Planck's length.
We will compute perturbations
 in the Einstein frame and then convert  the results back to the
original string frame for a physical interpretation.

In the Einstein frame the background equations for a generic
FRW universe are given by\footnote{Although we restrict our
attention to the case ${\cal K}=-1$, we will occasionally
keep  ${\cal K}$ in the
formulae for an easy comparison with the spatially-flat case.}:
\beq
{\cal H}'&=&-{1\over 6}\phi'^2,
\quad{\text{where}}\quad {\cal H}={{a'}\over a}\nn
{\cal H}^2+{\cal K}&=&{1\over 12}\phi'^2,\qquad\phi''+2{\cal H}\phi'=0\;,
\label{backeq}
\eeq
where a prime denotes differentiation with respect to the conformal
time $\eta$.
For ${\cal K}=-1$ the solution is just given by rewriting (\ref{Ssol})
 in the Einstein frame:
\beq
a(\eta)&=&\ell~(-\sinh\eta\cosh\eta)^{1\over 2}\nn
\phi(\eta)&=&-\sqrt 3~\ln(-\tanh\eta)+\phi_{\text{in}} \; , \; \;  
\eta < 0 \; ,
\label{esol}
\eeq
where the new modulus $\ell$, given by $\ell^2=L^2 \exp (\phi_{\text{today}}
 - \phi_{\text{in}})$,
replaces the string-frame classical modulus $L$.

To estimate quantum fluctuations around (\ref{esol}) we first go
over to isotropic spatial coordinates $(x,y,z)$ defined by
\beq
r=R\left(1+{{\cal K}\over 4}R^2\right)^{-1}\;,
\qquad{\text{where}}\quad R^2=x^2+y^2+z^2 \; ,
\eeq
and by the obvious identification of the angular coordinates.
In these coordinates the FRW metric takes the generic form
\beq
ds^2=a^2(\eta)\left(-d\eta^2+
\gamma_{ij}dx^idx^j\right),\qquad{\text{where}}\quad
\gamma_{ij}=\delta_{ij}\left(1+{{\cal K}\over  
4}R^2\right)^{-2},~i,j=1,2,3 \; ,
\eeq
and  generic perturbations are defined by
\beq
g_{\mu\nu}=g^{(0)}_{\mu\nu}+\delta g_{\mu\nu}\;\; , \;\;
\phi=\phi^{(0)}+\delta\phi \; ,
\eeq
where a superscript $(0)$ denotes the background solution.

We now consider the form of the action (\ref{Eaction})
 up to second-order terms in the fluctuations.
The calculations are long but straightforward.
After  using  the
 background equations
(\ref{backeq}), and after dropping
 irrelevant boundary terms (total divergences), the result can
be expressed covariantly in the form:
\beq
\FL
\delta^{(2)} S&=&{1\over 2\ell_P^2}\int d^4x\;\sqrt{-g}\,\bigg[
-{1\over 2}g^{\mu\nu}g^{\alpha\beta}g^{\lambda\sigma}\bigg(
\nabla_\lambda\delta g_{\beta\mu}\nabla_\sigma\delta g_{\nu\alpha}-
\nabla_\sigma\delta g_{\mu\nu}\nabla_\lambda\delta g_{\alpha\beta}\nn&+&
2\nabla_\alpha\delta g_{\mu\nu}\nabla_\sigma\delta g_{\beta\lambda}
-2\nabla_\lambda\delta g_{\beta\mu}\nabla_\nu\delta g_{\alpha\sigma}
\bigg)-g^{\mu\nu}\partial_\mu\delta\phi\partial_\nu\delta\phi
\nn&+&g^{\mu\nu}g^{\lambda\sigma}\partial_\lambda\phi
\delta\phi\nabla_\sigma\delta  
g_{\mu\nu}-2g^{\mu\lambda}g^{\nu\sigma}\partial_\lambda\phi\delta\phi
\nabla_\sigma\delta g_{\mu\nu}-2g^{\mu\lambda}g^{\nu\sigma}
\nabla_\sigma\partial_\lambda\phi\delta\phi\delta g_{\mu\nu}\;\bigg] \; ,
\label{secondorder}
\eeq
where, to this order, we can replace  $g_{\mu\nu}$ and $\phi$ by their
background expression (\ref{esol}), and all covariant derivatives are  
to be evaluated with respect to the background metric.

\section{Solving the perturbation equations}
\subsection{Tensor perturbations}

Since  tensor metric perturbations are automatically  
gauge-invariant, and decouple from   dilatonic
perturbations,  they are
easier to study.
They can be defined by
\beq
\delta g_{\mu\nu}^{\text {(T)}}={\text{diag}}(0,a^2 h_{ij})\;,
\label{tensor}
\eeq
where the symmetric three-tensor
$h_{ij}$ satisfies the transverse-traceless (TT) conditions
\beq
\nabla^i h_{ij}=0, \qquad {h^i}_i=0 \;,
\eeq
with $\nabla^{i}$ denoting the covariant derivative
with respect to $\gamma_{ij}$.
 Inserting (\ref{tensor}) into Eq. (\ref{secondorder}), and using  
(\ref{backeq}), we easily find:
\beq
\delta^{(2)} S^{(T)} = {1\over 4 \ell_P^2}\int d^4x\sqrt{\gamma}\; a^2\left(
{h'^{ij}}h'_{ij} - \nabla^l h^{ij}\nabla_l h_{ij}-2{\cal K}h^{ij}h_{ij}\right)
\; .
\label{tensoraction}
\eeq
For ${\cal K}=-1$,  tensor perturbations $h_{ij}$
can be expanded in TT  tensor 
pseudospherical harmonics  \cite{LKGS}
as
\beq
h_{ij}(\eta,{\bf x})=\int \! dn\;
\sum_{l=2}^{\infty} \sum_{m=-l}^{l}
h_{nlm}(\eta) (G_{ij}({\bf x}))^{n}_{lm}\; ,
\label{ttt}
\eeq
where the tensor harmonics
$(G_{ij})^{n}_{lm}$ satisfy the eigenvalue equation
\beq
\nabla^2(G_{ij}({{\bf x}}))^{n}_{lm}=-(n^2+3)(G_{ij}({{\bf
x}}))^{n}_{lm}\; .
\label{tensoreigen}
\eeq
Choosing their normalization so that:
\beq
\int d^3x \sqrt{h}(G_{ij}({\bf x}))^{n}_{lm}(G_{ij}({\bf x}))^{n'}_{l'm'}=
\delta(n-n')\delta_{ll'}\delta_{mm'}\; ,
\label{tensornorm}
\eeq
and inserting (\ref{ttt}) in (\ref{tensoraction}), we obtain
\beq
\delta^{(2)}S^{(T)}= {1\over 4 \ell_P^2}\int d\eta \;dn \;a^2 \sum_{l,m}
\left[(h_{nlm}')^2-(n^2+1)h_{nlm}^2\right].
\label{tensormode}
\eeq
Introducing finally the canonical variable
\beq
{u}_{nlm}=a  {h}_{nlm}\; ,
\label{varcanten}
\eeq
and using  the background equations    (\ref{backeq}), we get:
\beq
\delta^{(2)}S^{(T)}= {1\over 4 \ell_P^2}\int d\eta\; dn \; \; \sum_{l,m}
\left[({u}_{nlm}')^2-(n^2+{1\over 12}\phi'^2){u}_{nlm}^2\right],
\label{tensorcanon}
\eeq
yielding for ${u}_{nlm}$ the simple equation
\beq
{u}_{nlm}''+\left(n^2+{1\over 12}\phi'^2\right){u}_{nlm}=0\; .
\label{hmotion}
\eeq
Luckily, for the background (\ref{esol}),  Eq. (\ref{hmotion}) can be  
exactly solved
in terms of the standard hypergeometric function $F \equiv ~ _2F_1$  
\cite{ABST} by 
\beq
{u}_N(\eta)&=&C_1\;[{\text{csch}}^2(2\eta)]^{-{in\over 4}}F\,
\left[{1-in\over 4},{1-in\over 4},{2-in\over 2},-
{\text{csch}}^2(2\eta)\right]
\nn &+&C_2\;[{\text{csch}}^2(2\eta)]^{in\over 4}F\,\left[{1+in\over 4},
{1+in\over 4},{2+in\over 2},-{\text{csch}}^2(2\eta)\right]\; ,
\label{hypergeomten}
\eeq
where $N$ stands for the collection of indices $(nlm)$ and $C_{1,2}$ are
(classically arbitrary) integration constants.
In order to  correctly normalize the tensor perturbations, the action  
(\ref{tensorcanon}) has to be
quantized. At early times,
$n^2 \gg \phi'^2$, and thus $u$ is a free canonical field. Hence we
impose, as $\eta \rightarrow - \infty$,
\beq
{u}_{N}(\eta) \rightarrow
 {u}_{N}^{-\infty}(\eta) \equiv {2\ell_P \over \sqrt{n}}e^{-in\eta}.
\label{hearly}
\eeq
Using
$F[a,b,c,0]=1$, Eq. (\ref{hearly})  fixes the integration constants  
as $|C_1|=2\ell_P/\sqrt{n}, C_2 =0$.
The deviation from a trivial  plane-wave behaviour can easily be computed
 from the small argument limit of $F$. We find
\beq
u_N(\eta) = u_N^{-\infty}(\eta)\left(1+\alpha_n\,
e^{4\eta-i\beta_n}\right)\; ,
\label{earlyh}
\eeq
where  $\alpha_n,\beta_n$
are $n$-dependent constants fixed from the Taylor expansion of the
hypergeometric function. We note that the correction
to the vacuum amplitude dies off as
$e^{4\eta}$, i.e. as $t^{-4}$ in terms of cosmic time $t \sim -e^{-\eta}$.

We finally estimate the behaviour of the solution near the
singularity, i.e. for $\eta \rightarrow 0$, using \cite{ABST}
\beq
F\left[a,a,c,-{\text{csch}}^2(2\eta)\right]\simeq
{\Gamma(c)\over\Gamma(a)\Gamma(a+{1\over 2})}\left[
-2^{2a+1}|\eta|^{2a}\ln|\eta|\right]\; .
\eeq
Then, by virtue of the small $\eta$ behaviour
$a \simeq \ell|\eta|^{1/2}$ and of Eq. (\ref{varcanten}),  we find
\beq
|h_N|\simeq 2 \sqrt{{2 \over \pi }} {\ell_P \over \ell}
\sqrt{\coth{\left(n\pi\over 2
\right)}}\ln|\eta|\;.
\eeq
We shall come back to this result after deriving a similar
 expression for scalar perturbations.

\subsection{Scalar perturbations}

Consider now scalar metric-dilaton perturbations
defined by \cite{MFB}
\beq
\delta \phi ,\qquad
\delta g_{\mu\nu}^{\text {(S)}}=-a^2(\eta)\left(\matrix{2\varphi&\nabla_iB
\cr
                       \nabla_iB&2
(\psi\gamma_{ij}+\nabla_i\nabla_j E)\cr}\right)\; .
\label{scalarpert}
\eeq
 Inserting
(\ref{scalarpert}) in Eq. (\ref{secondorder}), and making use of
(\ref{backeq}), we find
\beq
\FL
\delta^{(2)} S^{(S)} &=&{1\over 2\ell_P^2}\int d^4x~a^2(\eta)\sqrt  
\gamma\;\bigg[(\delta
\phi')^2-\nabla\delta\phi\cdot\nabla\delta\phi+6\phi'\delta\phi \psi'-2
\varphi\phi'
\delta\phi'-2\phi'\delta\phi\nabla^2(B-E')\nn &-&12\psi'^2-8
\nabla \varphi\cdot\nabla
\psi+4(\nabla \psi)^2-24{\cal H}
\varphi\psi'+12{\cal K}(\varphi^2-\psi^2+2\varphi\psi)-
8\nabla \psi'\cdot\nabla B\nn 
&-&8{\cal H}\nabla \varphi\cdot\nabla B
-8{\cal H}\varphi\nabla^2E'-8\psi'\nabla^2E'+4{\cal K}(B-E')
\nabla^2(B-E')
\bigg]\; .
\label{secord}
\eeq
In (\ref{secord}) the variables $B,\varphi$ do not have  time derivatives
 and thus act as
 Lagrange multipliers, which provide constraints.
These  are:
\beq
0 = {\cal C_B}&\equiv &\phi'\delta\phi-4\psi'-4{\cal H}\varphi-4{\cal K}
(B-E') \nn
0 = {\cal C_\varphi}&\equiv&\phi'\delta\phi'-12{\cal K}\varphi+12{\cal H}
\psi'-4(\nabla^2+3{\cal K})\psi-
4{\cal H}\nabla^2(B-E')\; .
\label{constraints}
\eeq
 Following \cite{GMST}, we introduce the gauge-invariant
variable $\Psi$  by
\beq
\Psi={4\over\phi'}\big[\psi+{\cal H}(B-E')\big] \; ,
\label{gaugepsi}
\eeq
 and, after inserting the constraints, we recast the action  
(\ref{secord}) in the convenient form
\beq
\delta^{(2)}S^{(S)}={1\over 2\ell_P^2}\int d^4x\;a^2\sqrt \gamma
(\nabla^2+3{\cal K})\Psi\left[
\partial^2_\eta-\nabla^2+2({\cal H}'+{\cal K})\right]\Psi \; .
\label{scalaraction}
\eeq
 One can now make use of the constraints to eliminate the variable
$(B-E')$ from the action (\ref{scalaraction})
in terms of $\varphi,\psi$ and
$\delta\phi$.
The latter variables are
 not independent either, being related
by a linear combination of the two constraints $\cal C_\varphi,C_B$.
After its implementation the action (\ref{scalaraction}) contains only
true degrees of freedoms.

 In analogy with the case of tensor perturbations,
we introduce a canonical field $\Psi_c$ and expand it  as
\beq
\Psi_c \equiv a \Psi =
\int\!dn\sum_{l=0}^{\infty}\sum_{m=-l}^{l}\Psi_{nlm}
(\eta)Q_{nlm}({{\bf x}}),
\eeq
where $Q_{nlm}({{\bf x}})$ are the scalar pseudospherical harmonics,
satisfying \cite{LKGS}:
\beq
\nabla^2&&Q_{nlm}({{\bf x}})=-(n^2+1)Q_{nlm}({{\bf x}})\nn
\int d^3x&&\sqrt \gamma\;Q_{nlm}({{\bf x}})Q_{n'l'm'}({{\bf  
x}})=\delta(n-n')
\delta_{ll'}\delta_{mm'}\; .
\eeq
As a result, (\ref{scalaraction}) becomes
\beq
\delta^{(2)}S^{(S)}={1\over 2\ell_P^2}\int d\eta\;dn\left[(\bar\Psi'_N)^2-
(n^2- {1\over 4} \phi'^2)\bar\Psi_N^2\right]  , \quad N = (nlm) \; ,
\label{perturbaction}
\eeq
where  $\bar\Psi_N \equiv \sqrt{n^2+4}\Psi_N$.
The quantity $\bar\Psi_N$ enters the action in a canonical
way and therefore its vacuum fluctuations, like those of $u$, are easily
normalized. The equation for $\bar\Psi_N$ is simply
\beq
\bar\Psi_N''+ (n^2- {1\over 4} \phi'^2)\bar\Psi_N=0\; ,
\label{psibarmotion}
\eeq
 so that we must
impose, as $\eta \rightarrow - \infty$,
\beq
\bar\Psi_N(\eta) \rightarrow \bar\Psi_N^{-\infty}(\eta)&\equiv&{\ell_P\over \sqrt{n}}e^{-in\eta}\nn
\bar\Pi_N(\eta) \rightarrow \bar\Pi_N^{-\infty}(\eta)&\equiv&-i{\sqrt{n}\over \ell_P}e^{-in\eta}\; .
\label{normapsi}
\eeq
As  was the case for tensor perturbations, also Eq.
(\ref{psibarmotion})  can be transformed (for the background
(\ref{backeq})) into
 a hypergeometric equation. We find, specifically,
 \beq
\bar\Psi_N(\eta)&=&\tilde{C_1}\;[{\text{csch}}^2(2\eta)]^{-{in\over 4}}F\,
[{-1-in\over 4},{3-in\over 4},{2-in\over 2},-{\text{csch}}^2(2\eta)]
\nn &+&\tilde{C_2}\;[{\text{csch}}^2(2\eta)]^{in\over 4}F\,[{-1+in\over 4},
{3+in\over 4},{2+in\over 2},-{\text{csch}}^2(2\eta)]\; ,
\label{hypergeompsi}
\eeq
where, as before, we have to take
$|\tilde{C_1}|=\ell_P/\sqrt{n}, \tilde{C_2}=0$. Corrections to the free
 plane wave can be easily computed
and, again, are suppressed by four powers of $1/t$:
\beq
\bar\Psi_N(\eta)=\bar\Psi_N^{-\infty}(\eta)\left(1+\tilde\alpha_n\,
e^{4\eta-i\tilde\beta_n}\right)\; ,
\label{farpast}
\eeq
where $\bar\Psi_N^{-\infty}$ is given by (\ref{normapsi}) 
and
$\tilde\alpha_n,\tilde\beta_n$
are $n$-dependent constants fixed from the expansion of the
hypergeometric function.

To estimate the behaviour of (\ref{hypergeompsi}) near $\eta\simeq
0$, we use the formula \cite{ABST}
\beq
F\left[a,a+1,c,-{\text{csch}}^2(2\eta)\right]\simeq
{\Gamma(c)\over\Gamma(a+1)\Gamma(c-a)}\left[\right.&&\,
-2^{2a+3}a(a-c+1)|\eta|^{2(a+1)}\ln|\eta|\nn&&\left.+\,
2^{2a}|\eta|^{2a}\right]\; ,
\label{asymphyper}
\eeq
and obtain:
\beq
|\bar\Psi_N|\simeq \ell_P\sqrt{n^2+1\over 2\pi}\sqrt{\coth{\left(n\pi\over 2
\right)}}\left(-|\eta|^{3/2}\ln|\eta|+{2\over n^2+1}|\eta|^{-1/2}\right)\; .
\label{psinear0}
\eeq
\section{Discussion}
In order to discuss the physical significance of our
results it is useful to choose a convenient gauge.
In the spatially flat case it was found \cite{BGGMV} that the so-called
off-diagonal gauge \cite{Hwang,BGGMV}
was particularly useful in order to suppress the
large gauge artifacts present in the more commonly used \cite{MFB} longitudinal
gauge. The off-diagonal gauge is defined by setting
 $\psi=E=0$  in Eq. (\ref{scalarpert}). 
We shall now see how one can  
reconstruct  the scalar field fluctuation from
$\Psi$ in this gauge.

We first note that, in this gauge,
 the variables $\Psi$ and $B$ are related  through (\ref{gaugepsi})
as:
\beq
\Psi={4{\cal H}B\over\phi'}\; .
\label{relation}
\eeq
 Using Eq. (\ref{psibarmotion}) for $\bar\Psi_N$, as well as 
(\ref{relation}), we can derive the evolution equation for $B$:
\beq
B''-\nabla^2 B + \left(2{\cal H} - {4{\cal K} \over{\cal H}}\right)B' - 
( 4 {\cal H}^2+12{\cal K})B=0\; ,
\label{bequmotion}
\eeq
which agrees with Ref. \cite{BGGMV} for ${\cal K} = 0$. 
To relate $\delta\phi$ and $\Psi$ we first observe that
 the first of the two
constraints (\ref{constraints}) provides the relation
\beq
\phi'\delta\phi&=&4({\cal H}\varphi+{\cal K}B)\; ,
\label{rel3}
\eeq
while,  eliminating $\delta\phi$ from the two constraints 
(\ref{constraints}) and using (\ref{bequmotion}), we arrive at
a second relation 
\beq
\varphi&=&B'+2{\cal H}B \; .
\label{rel2}
\eeq
Combining  (\ref{rel2}) and (\ref{rel3}), and making use of 
(\ref{relation}), 
we are finally able to express $\delta\phi$ directly in terms
of $\Psi$ as 
\beq   
\delta\phi=\Psi'+{{\cal K}-{\cal H}'\over
{\cal H}}\Psi\; ,
\label{relphipsi}
\eeq
implying that $\delta\phi$ represents, in this gauge, a gauge-invariant
object.

It is  instructive to compare the ${\cal K}=-1$ case 
with  the spatially flat one, where the relevant 
gauge-invariant variable,  given by
\beq 
\psi^{\text{(gi)}}=\psi+{H\over\phi'}\delta\phi\; ,
\label{gaugeflat}
\eeq
becomes $\delta\phi$ itself in the off-diagonal gauge.
The {\it canonical field}, 
given by $v=a\delta\phi$, 
satisfies the well-known equation  
\cite{MFB}:
\beq
v''+\left(n^2-{z''\over z}\right)v=0,\qquad{\text{where}}\quad  
z={a\phi'\over
{\cal H}}.
\label{vmotion}
\eeq
Even in the
presence of spatial curvature,
the field $v$ still plays the role of the canonical
field in the far past, when $\eta$ is large and negative.
This can be checked by computing the equation
of motion for $v$ in the presence of curvature.
The explicit form of the equation
for $v$ is given by
\beq
v''+A_1v'+A_2v=0,\qquad A_1&=&-
{\cal K}^2\phi'^2\left[{\cal H}^3(n^2-{\cal K}+{3{\cal K}^2
\over {\cal H}^2})
\right]^{-1},\nn A_2&=&n^2+{{\phi'}^2 \over 12}
(1-{12{\cal K}\over {\cal H}^2})-({\cal H}
+{3{\cal K}\over {\cal H}})A_1 \;.
\eeq
Thus, as long as we are interested in the
early-time regime, $A_1$ is exponentially
small, $A_2 \rightarrow n^2$, and $v$
can be treated as the canonical field.

Using Eq. (\ref{relphipsi}), the behaviour  of $v$ in the far past
follows directly from that of $\bar\Psi_N$, given in Eqs.
(\ref{normapsi}), (\ref{farpast}):
\beq
v^{-\infty}(\eta)&\equiv&{\ell_P\over \sqrt{n}}\sqrt{2-in\over 2+in}
e^{-in\eta}\nn
\pi_v^{-\infty}(\eta)&\equiv&-i{\sqrt{n}\over \ell_P}\sqrt{2+in\over 2-in}
e^{-in\eta}\; ,
\label{vnormalization}
\eeq
with corrections again suppressed as $t^{-4}$, i.e.
\beq
v(\eta)=v^{-\infty}(\eta) \left(1+\hat\alpha_n\,
e^{4\eta-i\hat\beta_n}\right)\; ,
\label{earlyv}
\eeq
where ${\hat\alpha_n},\hat\beta_n$ are $n$-dependent constants.

We can  study how  other variables behave 
near $\eta \simeq 0$ by using their relation to $\Psi$ in this gauge
 and the behaviour of $\Psi$, Eq. 
(\ref{psinear0}). We easily find:
\beq
|B_N|\simeq{\ell_P\over\ell}\sqrt{n^2+1\over 2\pi}\sqrt{\coth({n\pi
\over 2})\over n^2+4}\left(-|\eta|\ln|\eta|+{2\over n^2+1}|\eta|^{-1}\right)\;,
\eeq
while
\beq
|\delta\phi_N|\simeq{\ell_P\over\ell}\sqrt{n^2+1\over 2\pi}
\sqrt{\coth({n\pi\over 2})\over n^2 +4}\ln|\eta|\; .
\eeq
 Let us finally compare the energy contained in the
quantum fluctuations of the dilaton and that in the classical solution 
near the singularity. Note that the expansion (\ref{asymphyper})
can be trusted
only up to some maximum $n$ for which $1\ll n_{\text{max}}\sim 1/|\eta|$.
Consequently, the ratio of the kinetic energy densities
near $|\eta|\simeq 0$ (up to constant prefactors of ${\cal O}(1)$) becomes
\beq
{{\cal E}_{\text Q}\over{\cal E}_{\text C}}=
{\int d^3x\sqrt{\gamma}\;
a^2(\delta\phi')^2
\over
\int d^3x\sqrt{\gamma}\;
a^2{\phi'}^2}
\simeq
{\ell_P^2 \over \ell^2} \int^{n_{\text{max}}}{dn\over n}\; n^3 \; .
\label{ratio}
\eeq
We can express the above result in terms of the value of
 the physical Hubble parameter
 $H(\eta)\equiv {\cal H}/a$ at horizon crossing
of the scale $n$, $H_{\text {HC}}(n)$, which is easily
computed as 
\beq
H_{\text {HC}}(n) \sim {1 \over \eta a}(\eta \sim 1/n) 
\sim n^{3/2}/\ell \;.
\label{HHC}
\eeq
Thus  (\ref{ratio}) takes the suggestive form
\beq
{{\cal E}_{\text Q}\over{\cal E}_{\text C}}= 
\ell_P^2  \int^{n_{\text{max}}}{dn\over n}\; H^2_{\text {HC}}(n) \; .
\label{horx}
\eeq

In general, in order to draw  physical conclusions, we should transform back
the results to the string frame. However, in our case, this is hardly
necessary. Concerning the importance of vacuum
 fluctuations as $\eta \rightarrow 0$,
we observe that the final result (\ref{horx}) expresses the relative
importance of quantum and classical fluctuations near the singularity in
terms of a frame-independent quantity, the ratio of the effective
Planck length to the size of the horizon. Since, by definition of the
perturbative dilaton phase, the Hubble radius is always larger 
than the string scale,
we find that the relative importance of quantum fluctuations is always
bounded by the ratio $\ell_P / \ell_s$ which is always
less than one in the perturbative phase.

 Let us now come to the more subtle issue of
the far-past behaviour of tensor and
scalar quantum fluctuations. 
Computations may be done in either frame,
since the dilaton is approximately constant in the far past. Our results,
expressed in Eqs. (\ref{earlyh}) and (\ref{earlyv}), show
 that corrections to the trivial quantum fluctuations are of relative order
 $e^{4\eta} \sim t^{-4}$, i.e. of order $t^{-3}$ relative to the 
(homogeneous) classical 
 perturbation. This suggests that quantum effects do not
modify appreciably classical behaviour in the far past,
 in contrast to the claim
made in \cite{KLB}.
This attitude is also supported 
 by the structure of the superstring one-loop effective-action 
 (which is well-defined thanks to the string  cutoff). 
Because of supersymmetry,
 neither a cosmological term nor a renormalization of Newton's constant
are generated at one-loop, but only
 terms containing at least four derivatives. As a result, quantum corrections
to early-time classical behaviour are of relative order $t^{-6}$, 
i.e just like our corrections $(\delta\phi'/\phi')^2$. Note, incidentally,
that generating a cosmological constant by quantum corrections would upset
completely the whole PBB scenario.

  We also see, however, that, as claimed in \cite{KLB},
the leading (free-theory)
fluctuations (the $1$'s in Eqs. (\ref{earlyh}) and (\ref{earlyv})) dominate
over the homogeneous classical perturbation  by one power of $t$.
If taken at face value, they
upset classical behaviour at early-enough times,
 $|t| > \ell^2/\ell_P$ \cite{KLB}.
  The answer to the issue
raised in \cite{KLB} thus appears to depend
on whether  (zero-point, non-amplified) vacuum quantum fluctuations
in the (trivial) Milne background can give
physically important effects on the scale of Milne's Hubble radius $H^{-1}
 \sim t$. A complete clarification of 
this point would be certainly desirable. 

We stress however that, irrespectively of the final answer
to this issue, vacuum fluctuations have the same
time dependence as the  typical
 {\it inhomogeneous} classical perturbation  discussed 
in \cite{BMUV,BDV}, but much smaller
amplitudes. Indeed,  
 an initial classical state apt to give rise to a pre-big bang event
(i.e. to gravitational collapse in the Einstein frame) 
in a  region of space of size $\ell_{in}\gg \ell_P$
must correspond, quantum mechanically, to having
 parametrically large occupation numbers in certain
quantum states \cite{BDV}. Such a quasi-classical 
configuration  cannot be appreciably affected by quantum fluctuations
${\cal O}(1)$ in those occupation numbers.

\acknowledgments
We wish to thank Maurizio Gasperini, Massimo Giovannini and Slava  
Mukhanov for helpful discussions. We also thank 
Nemanja Kaloper and Andrei Linde for discussions and correspondence 
 which helped understanding the relation of this work to theirs.
The work of AG was supported in part by World Laboratory.
The work of GP was supported in part by the Angelo Della Riccia  
Foundation.


\end{document}